\documentclass[a4paper,11pt]{article}
\usepackage{pos}
\usepackage{bbold}
\usepackage{graphicx} 
\usepackage{booktabs} 
\usepackage{epstopdf}
\usepackage{slashed}
\usepackage{physics}
\usepackage{simpler-wick}
\usepackage[absolute,overlay]{textpos}
\title{Prediction and compression of lattice QCD data using machine learning algorithms on quantum annealer}
\ShortTitle{Prediction and compression of lattice QCD data using ML algorithms on quantum annealer}

\author*[a]{Boram Yoon}
\author[b,c,d]{Chia Cheng Chang}
\author[e]{Garrett T. Kenyon}
\author[e]{Nga T.T. Nguyen}
\author[d]{Ermal Rrapaj}

\affiliation[a]{CCS-7, Computer, Computational and Statistical Sciences Division, Los Alamos National Laboratory,Los Alamos, NM 87545, USA}

\affiliation[b]{RIKEN iTHEMS, Wako, Saitama 351-0198, Japan}

\affiliation[c]{Department of Physics, University of California, Berkeley, California 94720, USA}

\affiliation[d]{Nuclear Science Division, Lawrence Berkeley National Laboratory, Berkeley, California 94720, USA}

\affiliation[e]{CCS-3, Computer, Computational and Statistical Sciences Division, Los Alamos National Laboratory,Los Alamos, NM 87545, USA}
  
\emailAdd{boram@lanl.gov}

\abstract{
We present regression and compression algorithms for lattice QCD data utilizing the efficient binary optimization ability of quantum annealers. In the regression algorithm, we encode the correlation between the input and output variables into a sparse coding machine learning algorithm. The trained correlation pattern is used to predict lattice QCD observables of unseen lattice configurations from other observables measured on the lattice. In the compression algorithm, we define a mapping from lattice QCD data of floating-point numbers to the binary coefficients that closely reconstruct the input data from a set of basis vectors. Since the reconstruction is not exact, the mapping defines a lossy compression, but, a reasonably small number of binary coefficients are able to reconstruct the input vector of lattice QCD data with the reconstruction error much smaller than the statistical fluctuation. In both applications, we use D-Wave quantum annealers to solve the NP-hard binary optimization problems of the machine learning algorithms.
}

\FullConference{%
 The 38th International Symposium on Lattice Field Theory, LATTICE2021
  26th-30th July, 2021
  Zoom/Gather@Massachusetts Institute of Technology
}

\begin{document}
\maketitle

\section{Introduction}
In lattice quantum chromodynamics (QCD) simulations, a large number of observables are measured on each Monte Carlo sample of the gauge field. The modern lattice QCD calculations targeting precise estimates often require the observable measurements computationally more expensive than the generation of the gauge field and produce $\mathcal{O}$(PB) of data that need to be stored for a long term analysis. Since the measured observables share the same background lattice, however, their statistical fluctuations are correlated with each other. By exploiting the correlations, one can predict the values of the computationally expensive observables from the values of the computationally cheap observables, and compress the data for significantly reduced storage requirements.

Machine learning (ML) techniques provide powerful algorithms building models based on the correlation pattern of data. Various approaches enhancing the lattice QCD calculations using ML have been explored in recent studies~\cite{Yoon:2018krb, Zhang:2019qiq, Kanwar:2020xzo, Bulusu:2021rqz}. However, the ML training, which is the procedure of finding optimal model parameters describing the data, of most of the ML algorithms involve nontrivial optimization problems. The optimization problems are computationally challenging as they are typically non-convex and the dimension of optimization parameters is very large. Stochastic approaches, such as the Adam optimizer~\cite{kingma2017adam}, are used in many algorithms, but it demands expensive computations to obtain a near-optimal solution, especially for large models.

The optimization problems in ML algorithms can be solved using quantum annealers, such as the D-Wave system~\cite{dwave_system}, which realizes the Ising spin system at a very low temperature. Starting from a transverse field Hamiltonian, the quantum processor of the D-Wave system performs quantum annealing towards the target Ising Hamiltonian. After the quantum annealing, the final state of the Ising spin system falls into a low-lying energy solution of the Ising Hamiltonian. Since the annealing time for each sample takes only $O(10)$ microseconds, a solution close to the ground-state can be obtained within a short wallclock time by repeating the quantum annealing and taking the smallest energy solution. Therefore, the optimization problem in ML algorithms can be solved using quantum annealing by mapping the optimization problem  onto the structure of the Ising Hamiltonian:\looseness-1
\begin{equation}
	H(\boldsymbol{h}, \boldsymbol{J}, \boldsymbol{s}) = \sum_i^{N_q} {h_i s_i} + \sum_{i<j}^{N_q} {J_{ij}  s_i s_j }\,,
	\label{eq:H_Dwave}
\end{equation}
where $s_{i}=\pm1$ is the binary spin variable, and $h_i$ and $J_{ij}$ are real-valued qubit biases and coupling strengths that can be set by a user. Note that this Ising model is isomorphic to a quadratic unconstrained binary optimization (QUBO) problem when defined in terms of $a_i = (s_i+1)/2$.

Utilizing quantum annealing, in this work, we present a ML regression algorithm predicting the values of the lattice QCD observables~\cite{Nguyen:2019gpo} and a lossy data compression algorithm reducing the data storage requirement of the lattice QCD data~\cite{Yoon:2021btl}.

\section{Regression algorithm using quantum annealer}
ML prediction of the unmeasured but expensive observables from the measured but cheap observables was first discussed in Ref.~\cite{Yoon:2018krb} and applied to a prediction of quasiparton distribution function matrix elements in Ref.~\cite{Zhang:2019qiq}. The basic idea is the following. Assume that we have measurements on $M$ independent lattice configurations. A set of common observables $\mathbf{X}$ are measured on all $M$ configurations, while a target observable $O$ is measured only on the first $N\ll M$ configurations. We call the first $N$ configurations the \emph{labeled dataset}, and the rest of the $M-N$ configurations the \emph{unlabeled dataset}. Using the labeled dataset, one can train a ML regression model $f$ to yield the prediction of $O$ from $\mathbf{X}$ for each configuration as $O^{(k)}_P = f(\mathbf{X}^{(k)})$, where $k$ is the configuration index, and $O^{(k)}_P$ is the ML prediction of $O^{(k)}$. After training the correlation pattern between $\mathbf{X}$ and $O$, the trained model can make predictions of $O^{(k)}$ for the unlabeled dataset.

Quantification of the prediction uncertainties is an important issue when applying ML to scientific data. All ML predictions may have prediction errors since the trained model is empirical, and the data used for the training have statistical noise. When estimating the expectation value of an observable through the Monte Carlo average, therefore, the simple average over the machine learning predictions $\frac1M\sum_k O^{(k)}_P$ could be a biased estimator of the target expectation value $\langle O \rangle$. For ML predictions on statistical data, however, an unbiased estimator can be defined following the bias correction algorithms developed in the truncated solver method~\cite{Bali:2009hu} and the all-mode averaging~\cite{Blum:2012uh}. In this approach, the labeled dataset is split into training ($N_t$) and bias-correction ($N_b$) datasets, where $N_t+N_b = N$, and the bias correction of the regression model trained on the training dataset can be carried out using the bias correction dataset as following:
\begin{align}
  O^{BC}_P = \frac{1}{M}\sum_{k=N+1}^{N+M} O^{(k)}_P + \frac{1}{N_b}\sum_{k=N_t+1}^{N} \left(O^{(k)} - O^{(k)}_P\right)\,,
  \label{eq:unbiased}
\end{align}
where $O^{BC}_P$ is an unbiased estimator of the expectation value of $O$: $\langle O^{BC}_P \rangle = \langle O_{P} \rangle + \langle O - O_{P} \rangle = \langle O \rangle$. Here the second term is the bias correction term estimating the difference between the ground truth and its ML prediction.  In the calculation of the statistical error of $O^{BC}_P$, the bias correction term naturally increases the size of the statistical error accounting for the prediction quality, so it provides a statistically sound uncertainty quantification of the ML predictions.

We design a new regression algorithm based on sparse coding~\cite{rozell.06c}, which is an unsupervised machine learning technique. For a set of input vectors of $\{\mathbf{X}^{(k)}\in\mathbb{R}^D\}_{k=1}^M$, sparse coding finds a matrix $\boldsymbol{\phi}\in\mathbb{R}^{D\times N_q}$, which is called the sparse coding dictionary, and the minimum number of nonzero coefficients $\{\boldsymbol{a}^{(k)}\in\mathbb{R}^{N_q}\}_{k=1}^M$ such that they reconstruct the input data as the following linear equation: $\mathbf{X}^{(k)} \approx \boldsymbol{\phi}\boldsymbol{a}^{(k)}$. This can be written as the following optimization problem:
\begin{align}
	\min\limits_{ \boldsymbol{\phi} }\sum_{k}\min\limits_{ \boldsymbol{a}^{(k)} } \left[  \, \frac{1}{2}  ||  \mathbf{X}^{(k)} - \boldsymbol{\phi} \boldsymbol{a}^{(k)} ||^2	+	\lambda || \boldsymbol{a}^{(k)} ||_0 \, \right]\,,
	\label{eq:H_SC}
\end{align}
where $\lambda$ is the sparsity penalty parameter. Here the difficult part is the optimization in $\boldsymbol{a}^{(k)}$ because the $L_0$-norm in the sparsity penalty term makes the problem NP-hard~\cite{Natarajan:1995:SAS:207985.207987}.

The $\boldsymbol{a}^{(k)}$-optimization part can be mapped onto the D-Wave through the following~\cite{Nguyen:2016,Nguyen2018ImageCU}:
\begin{equation}
	\boldsymbol{h} = -\boldsymbol{\phi}^{T}  \mathbf{X} + \lambda + \frac{1}{2}\,, \qquad
	\boldsymbol{J} 	= \boldsymbol{\phi}^{T} \boldsymbol{\phi}\,, \qquad
	\boldsymbol{s} = 2\boldsymbol{a}-1\,.
\label{eq:hQ}
\end{equation}
In this way, one can solve the optimization in $\boldsymbol{a}^{(k)}$ using D-Wave, while the optimization for $\boldsymbol{\phi}$ can be done on classical computers. Note that, on the D-Wave, $a_i$ corresponds to the spin up or down state, so it is restricted to a binary variable: $\boldsymbol{a}^{(k)}\in \{0,1\}^{N_q}$.

Once trained, the $\boldsymbol{\phi}$ encodes the correlation pattern of the input data $\mathbf{X}^{(k)}$. Assume that we have a new vector $\mathbf{X}'$ that has a few corrupted vector elements. If one calculates the sparse coefficients $\boldsymbol{a}'$ of the $\mathbf{X}'$ using the trained $\boldsymbol{\phi}$, it finds the most plausible basis vectors among $\boldsymbol{\phi}$ describing the uncorrupted input vector elements. In the reconstruction $\boldsymbol{\phi}\boldsymbol{a}'$, as a result, the corrupted input vector elements are replaced by those considered more natural based on the trained correlation pattern of $\boldsymbol{\phi}$. In the field of image processing, this procedure is called \emph{inpainting} or \emph{denoising}~\cite{4959679}.

Using the inpainting feature of the sparse coding, a regression model can be constructed as follows. Consider $N$ sets of labeled data $\{\mathbf{X}^{(i)}, y^{(i)}\}_{i=1}^N$, and $M$ sets of the unlabeled data $\{\mathbf{X}^{(j)}\}_{j=N+1}^{N+M}$, where $\mathbf{X}^{(i)}\equiv\{x_1^{(i)}, x_2^{(i)}, \ldots, x_D^{(i)}\}$ is the input vector (independent variable), and $y^{(i)}$ is the output variable (dependent variable). The goal is to build a regression model $F$ using the labeled dataset to make predictions $\hat{y}$ of $y$ for an unseen input data $\mathbf{X}$ as $F(\mathbf{X}) = \hat{y} \approx y$. The key idea is to concatenate the input and output variables of the labeled dataset into an extended vector, $\widetilde{\mathbf{X}}^{(i)}\equiv\{x_1^{(i)}, x_2^{(i)}, \ldots, x_D^{(i)}, y^{(i)}\}$, and calculate the dictionary matrix $\widetilde{\boldsymbol{\phi}} \in \mathbb{R}^{(D+1)\times N_q}$ of the extended vectors $\{\widetilde{\mathbf{X}}^{(i)}\}_{i=1}^{N}$. For the prediction on the unlabeled dataset, build vectors $\widetilde{\mathbf{X}}_o^{(j)}\equiv\{x_1^{(j)}, x_2^{(j)}, \ldots, x_D^{(j)}, \bar{y}^{(j)}\}$, where $\bar{y}^{(j)}$ is an initial guess of $y^{(j)}$ set by the average value of $y^{(i)}$ in the labeled dataset.  Then, using the dictionary $\widetilde{\boldsymbol{\phi}}$ obtained from the labeled dataset, find sparse representation $\boldsymbol{a}^{(j)}$ for $\widetilde{\mathbf{X}}^{(j)}_o$ and calculate reconstruction as $\widetilde{\mathbf{X}}'^{(j)} = \widetilde{\boldsymbol{\phi}}\boldsymbol{a}^{(j)}$. This replaces the outlier components, including $\bar{y}^{(j)}$, in $\widetilde{\mathbf{X}}^{(j)}_o$ by the values that can be described by $\widetilde{\boldsymbol{\phi}}$. Finally, the prediction of $y^{(j)}$ is given by $\hat{y}^{(j)} =  [\widetilde{\mathbf{X}}'^{(j)}]_{D+1}$. A detailed procedure, including the pre-training and normalization improving the prediction performance, is described in Ref.~\cite{Nguyen:2019gpo}.

\begin{figure}[tb]
\centering
  	\includegraphics[width=0.5\textwidth]{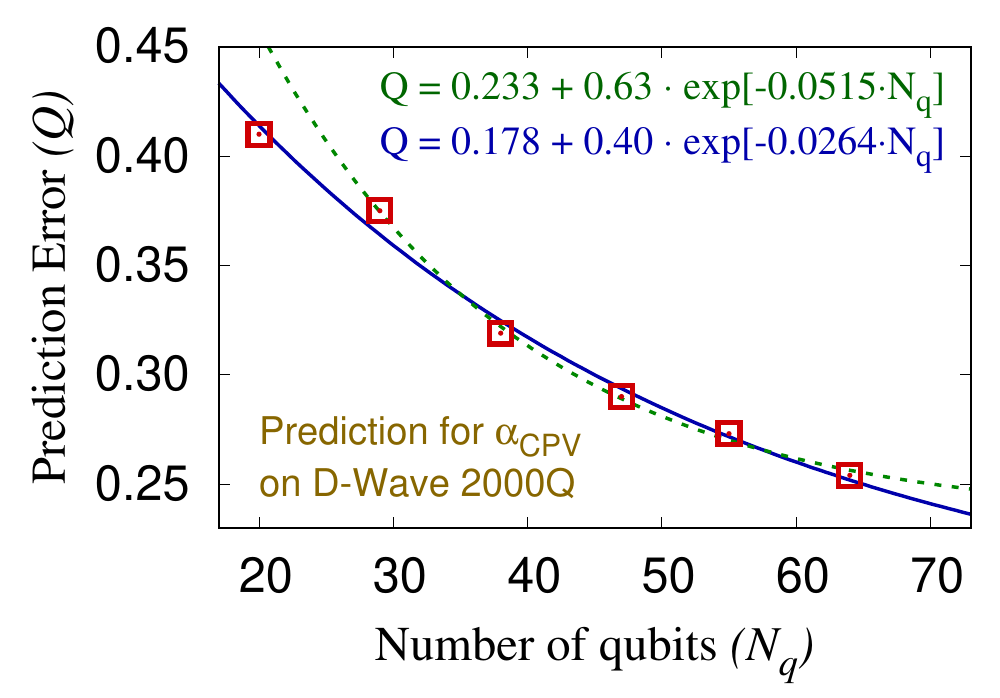}
    \caption{Prediction error of the CP-violating two-point correlation functions predicted using the sparse coding regression algorithm as a function of the number of qubits used for the prediction (red squares). An exponential ansatz is fitted to the data points for all $N_q$ (blue line) and those excluding $N_q=20$ (green line).
  }
  \label{fig:std_Nq}
\end{figure}

To demonstrate the regression ability of the proposed algorithm, we apply it to the lattice QCD simulation data generated for the calculation of the charge-parity (CP) symmetry violating phase $\alpha_{\textrm{CPV}}$ of the neutron~\cite{Yoon:2017tag}. From the real and imaginary parts of the normal and pseudoscalar-projected two-point correlation functions calculated without CPV interactions at $t/a=8, 9, \ldots, 12$, we predict the imaginary part of the two-point correlation functions calculated in presence of the CP-violating interactions. In this example, the input vector has 20 variables and the output is a single variable. For this test, we train $\widetilde{\boldsymbol{\phi}}$ using 6976 data samples and test its prediction ability on 8640 data samples measured in Refs.~\cite{Bhattacharya:2016oqm,Bhattacharya:2016rrc}. For the training and prediction, the $\boldsymbol{a}$-optimization is carried out using the D-Wave 2000Q quantum annealer, and the $\boldsymbol{\phi}$-optimization is performed using the stochastic gradient descent method on a classical computer. Because of the restricted number of fully connected logical qubits on the D-Wave 2000Q hardware, the maximum dimension of the sparse representation $\boldsymbol{a}^{(k)}$ is limited to 64. Results presented in Fig.~\ref{fig:std_Nq} show that the current prediction performance is limited by the maximum number of qubits of the quantum annealer. A detailed description of the quantum annealing and regression results are presented in  Ref.~\cite{Nguyen:2019gpo}.

\section{Lossy compression algorithm using quantum annealer}

Another interesting use of quantum annealers is data compression. Lattice QCD simulations produce a huge amount of data, but they are correlated with each other, so one can reduce the storage space requirement by exploiting the correlation. In most lattice QCD data analyses, furthermore, data precision sufficiently smaller than the statistical fluctuation is good enough, so lossy data compression is considered as a viable approach. 

A lossy compression algorithm can be constructed in a similar structure as the sparse coding described in the previous section. Assume that we have a matrix $\boldsymbol{\phi}\in \mathbb{R}^{D\times N_q}$ and binary coefficients $\boldsymbol{a}^{(k)}\in \{0,1\}^{N_q}$ that precisely reconstruct the input vectors $\mathbf{X}^{(k)}\in \mathbb{R}^{D}$ such that $\mathbf{X}^{(k)} \approx \boldsymbol{\phi} \boldsymbol{a}^{(k)}$ for all data index $k=1,2,3,\ldots, N$. The procedure defines a mapping from $\mathbf{X}$-space to $\boldsymbol{a}$-space:
\begin{align}
\left\{\mathbf{X}^{(k)}\vert\mathbf{X}^{(k)}\in \mathbb{R}^{D}\right\}_{k=1}^N
\longrightarrow 
\left(\left\{\boldsymbol{a}^{(k)} \vert \boldsymbol{a}^{(k)}\in \{0,1\}^{N_q}\right\}_{k=1}^N, \boldsymbol{\phi}\in \mathbb{R}^{D\times N_q}\right)\,.
\end{align}
Note that here we restrict the $a^{(k)}_i$ to a binary variable so that it can be stored in a single bit. Also, we focus on the cases where 
$N_q \ll N$, which is the typical situation in high-statistics data, so that the memory usage of $\boldsymbol{\phi}$ is sufficiently smaller than the size of the original data. This transformation forms a data compression since the data in $\boldsymbol{a}$-space require less memory than those in $\mathbf{X}$-space. 

Such $\{\boldsymbol{a}^{(k)}\}$ and $\boldsymbol{\phi}$ can be obtained by minimizing the mean square reconstruction error,
\begin{align}
	\min\limits_{ \boldsymbol{\phi} }\sum_{k}\min\limits_{ \boldsymbol{a}^{(k)} } \left[  \,   \sum_{i=1}^D 
	\left( X^{(k)}_i - \big[\boldsymbol{\phi} \boldsymbol{a}^{(k)}\big]_i \right)^2 \, \right]\,.
	\label{eq:E_SC}
\end{align}
In general, a weight factor of inverse variance, $1/\sigma_{X_i}^2$, should be multiplied to the least-squares loss function to avoid the algorithm focusing on the large-variance components of the input vector. In this study, we assume that the data is standardized before the compression, so $1/\sigma_{X_i}^2=1$. Here, the $\boldsymbol{\phi}$-optimization part can be performed on classical computers using stochastic gradient descent, but finding the optimal solution of the $\boldsymbol{a}$-vectors is an NP-hard problem because it is a binary optimization. However, the $\boldsymbol{a}$-optimization can be solved using the D-Wave quantum annealer after the transformation to the Ising Hamiltonian in Eq.~\eqref{eq:H_Dwave} following
\begin{align}
\mathbf{J} = 2\boldsymbol{\phi}^T\boldsymbol{\phi}, \qquad {h}_i = -2\left[\boldsymbol{\phi}^T \mathbf{X}\right]_i + \left[\boldsymbol{\phi}^T\boldsymbol{\phi}\right]_i, \qquad \boldsymbol{s} = 2\boldsymbol{a} -1\,.
\label{eq:QUBO-transf}
\end{align}

Since this is a lossy compression algorithm, the reconstruction is not exactly the same as the original data, and there could be reconstruction errors that introduce bias in the data analysis. However, the bias can be removed using a small portion of the original data following the bias correction approach given in Eq.~\eqref{eq:unbiased}. A detailed description can be found in Ref.~\cite{Yoon:2021btl}.

\begin{figure}[tb]
    \centering
    \includegraphics[width=0.45\textwidth]{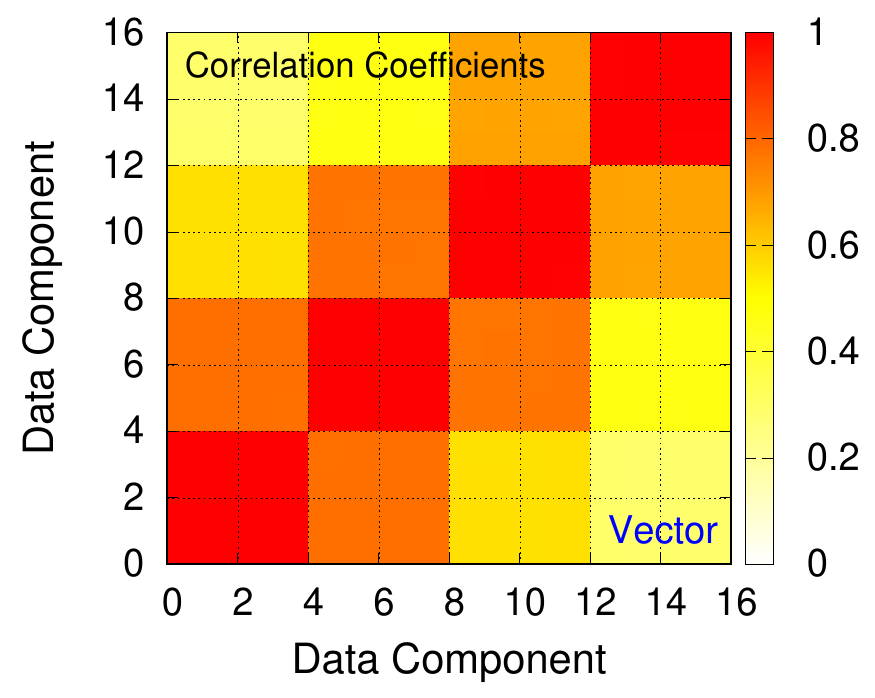} \qquad
    \includegraphics[width=0.45\textwidth]{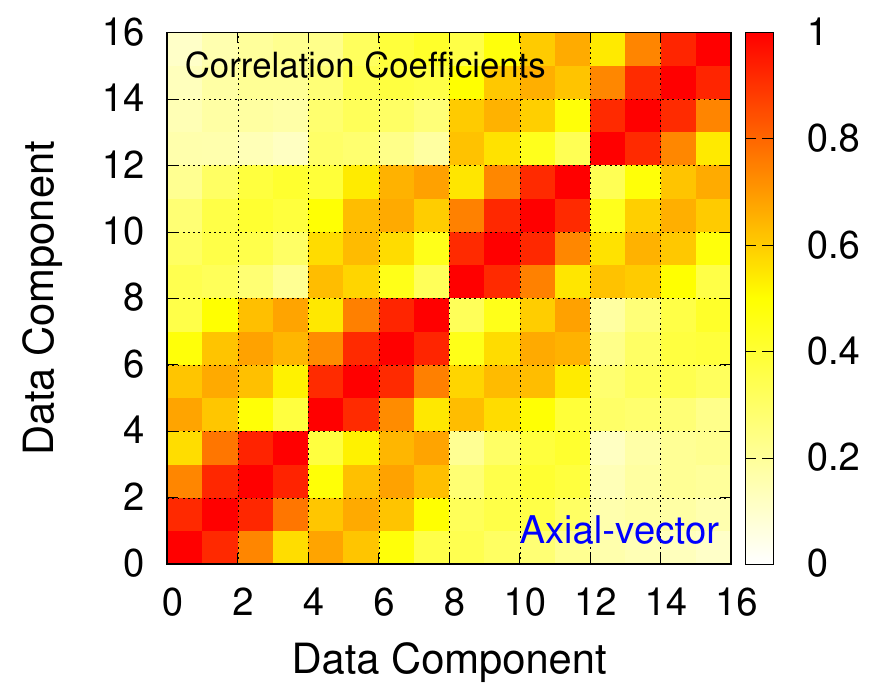}
    \caption{Correlation pattern of the 16 components of the vector (left) and the axial-vector (right) data. Red indicates the high correlation (correlation coefficient = 1), and white indicates no correlation.}
    \label{fig:data}
\end{figure}

\begin{figure}[tb]
    \centering
    \includegraphics[width=0.45\textwidth]{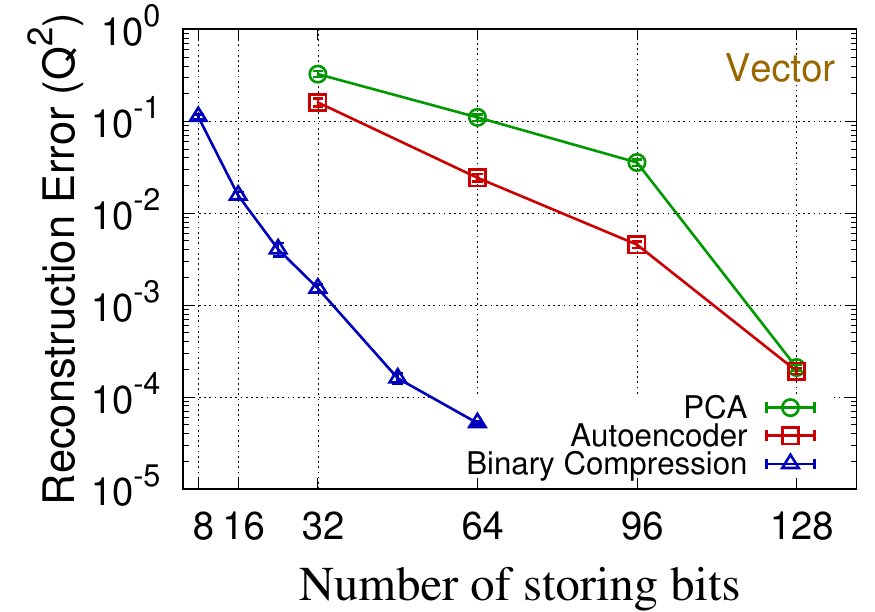}\qquad
    \includegraphics[width=0.45\textwidth]{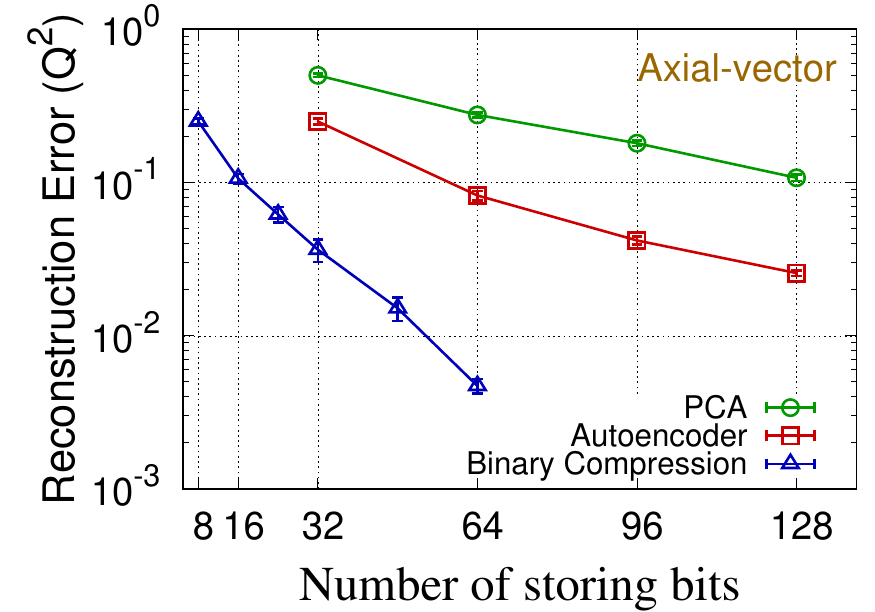}
    \caption{Squared reconstruction error for different number of storing bits. For the principal component analysis (PCA) and autoencoder (AE) approaches, the number of storing bits is calculated by $32\times N_z$, assuming single-precision floating-point numbers.}
    \label{fig:Q-simann}
\end{figure}

To show the compression performance, we apply the algorithm to compress the vector and axial-vector nucleon three-point correlator data of the 16 timeslices calculated in Ref.~\cite{Gupta:2018zkh}. Fig.~\ref{fig:data} shows their correlation pattern. To avoid the effects from the integrated control errors (ICE)~\citep{dwave_ice} of the D-Wave hardware, we first carry out the compression tests using D-Wave's simulated annealing sampler implemented in the D-Wave's Ocean library~\cite{dwave_ocean}, on classical computers. Fig.~\ref{fig:Q-simann} shows the reconstruction error of the new algorithm in blue data points, and those of the other compression algorithms based on principal component analysis (PCA) and neural-network autoencoder for various different numbers of storing bits. The new binary compression algorithm shows a much smaller reconstruction error than others. Also, by comparing the vector and axial-vector results, one can see that the reconstruction error of the vector data is much smaller than the reconstruction error of the axial-vector data, because of the higher correlation in the vector channel. We also carry out the compression test using D-Wave 2000Q and Advantage systems but found slightly worse compression performance than the simulated annealing due to the ICE. 

The compression can be used for outlier detection for rare events or data corruption, because a large reconstruction error for an input data vector indicates that the correlation pattern of the data is different from those used in the training of $\boldsymbol{\phi}$. Also, the relation between the original vector $\mathbf{X}$ and the binary coefficients $\boldsymbol{a}$ is linear, so one can replace the expensive operations on the floating-point numbers of $\mathbf{X}$ by the cheap operations on the single-bit coefficients $\boldsymbol{a}$.

\section{Conclusion}
We presented a new ML regression algorithm utilizing D-Wave quantum annealer as an efficient optimization algorithm finding optimal ML model parameters. Motivated by the inpainting feature of the sparse coding, a correlation pattern between the dependent and independent variables is trained to make predictions on unlabeled dataset. For the training and predictions, the D-Wave 2000Q quantum annealer is used to find the sparse coefficients of the sparse coding ML algorithm. The regression algorithm is applied to predict the two-point correlators of the CP-violating interactions. Results show that the regression performance is limited by the number of fully connected logical qubits of current D-Wave hardware, but promising on the future quantum annealing architectures.

We also presented a new lossy data compression algorithm based on ML using a D-Wave quantum annealer. The algorithm finds a set of basis vectors and their binary coefficients that precisely reconstruct the original data. Finding such binary coefficients is an NP-hard optimization problem, but the algorithm is formulated so that they can be obtained using the D-Wave quantum annealer with a short wallclock time. The compression algorithm applied to nucleon three-point correlator data shows a much better compression performance than the compression algorithms based on PCA or the autoencoder. Currently, the compression performance with D-Wave quantum annealer is limited by the ICE of the D-Wave hardware. On future architectures with improved ICE, however, the algorithm is expected to perform much better.

\section*{Acknowledgments}
The QUBO optimizations were carried out using the D-Wave 2000Q at Los Alamos National Laboratory (LANL) and the D-Wave's Leap Quantum Cloud Service. Simulation data used for the numerical experiment were generated using the computer facilities at NERSC, OLCF, USQCD Collaboration, and institutional computing at LANL. This work was supported by the U.S. Department of Energy (DOE), Office of Science, Office of High Energy Physics under Contract No. 89233218CNA000001 and LANL LDRD program. Lawrence Berkeley National Laboratory (LBNL) is operated by The Regents of the University of California (UC) for the U.S. DOE under Federal Prime Agreement DE-AC02-05CH11231. This material is based upon work supported by the U.S. Department of Energy, Office of Science, Office of Nuclear Physics, Quantum Horizons: QIS Research and Innovation for Nuclear Science under Award Number FWP-NQISCCAWL (CCC). ER acknowledges the NSF N3AS Physics Frontier Center, NSF Grant No. PHY-2020275, and the Heising-Simons Foundation (2017-228).


\bibliographystyle{JHEP}
\bibliography{ref}
\end{document}